\documentclass[conference]{IEEEtran}

\usepackage[british]{babel}
\usepackage{graphicx}
\usepackage[hyphens]{url}
\usepackage{enumerate}
\usepackage[noadjust]{cite}
\usepackage[pdftex,colorlinks=true]{hyperref}

\begin{document}
%
\title{Reproducibility as a Technical Specification}

\author{\IEEEauthorblockN{Tom Crick}
\IEEEauthorblockA{Department of Computing \& Information Systems\\
Cardiff Metropolitan University, UK\\
{\url{tcrick@cardiffmet.ac.uk}}}
\and
\IEEEauthorblockN{Benjamin A. Hall}
\IEEEauthorblockA{MRC Cancer Unit\\
University of Cambridge, UK\\
{\url{bh418@mrc-cu.cam.ac.uk}}}
\and
\IEEEauthorblockN{Samin Ishtiaq}
\IEEEauthorblockA{Microsoft Research\\
Cambridge, UK\\
{\url{samin.ishtiaq@microsoft.com}}}}



\maketitle

\begin{abstract}
Reproducibility of computationally-derived scientific discoveries
should be a certainty. As the product of several person-years' worth
of effort, results -- whether disseminated through academic journals,
conferences or exploited through commercial ventures -- should at some
level be expected to be repeatable by other researchers. While this
stance may appear to be obvious and trivial, a variety of factors
often stand in the way of making it commonplace. Whilst there has been
detailed cross-disciplinary discussions of the various social,
cultural and ideological drivers and (potential) solutions, one factor
which has had less focus is the concept of reproducibility as a
\emph{technical} challenge. Specifically, that the definition of an
\emph{unambiguous} and \emph{measurable} standard of reproducibility
would offer a significant benefit to the wider computational science
community.

In this paper, we propose a high-level technical specification for a
service for reproducibility, presenting cyberinfrastructure and
associated workflow for a service which would enable such a
specification to be verified and validated. In addition to addressing
a pressing need for the scientific community, we further speculate on
the potential contribution to the wider software development community
of services which automate \emph{de novo} compilation and testing of
code from source. We illustrate our proposed specification and
workflow by using the {\emph{BioModelAnalyzer}} tool as a running
example.
\end{abstract}

%


\section{Introduction}

This paper aims to start a discussion in the wider computational
science community about the reproducibility of its algorithms, models,
tools and benchmarks. There is a significant opportunity for this
(broad and diverse) community --- of whom we are proud members --- to
identify and address the technical and socio-cultural issues
surrounding reproducibility in both their specific research domain as
well as more broadly for computational science; a desirable outcome
would be a clear specification to encourage, enable and ultimately
enforce reproducibility. Enumerating a standard for reproducibility
would have a clear benefit for researchers as well as the wider
community as a whole.

Over the past decade, we have seen a step-change in how science and
engineering is done. Experiments, simulations, models, benchmarks,
even proofs cannot be done without leveraging software and
computation. A 2012 report by the Royal Society summarised that
computational techniques have ``{\emph{moved on from assisting
scientists in doing science, to transforming both how science is done
and what science is done}}''~\cite{rssaaoe:2012}, potentially adding
another ``pillar'' to the scientific
method~\cite{hey:2009,vardi-cacm-2010}. Thus, the reproduction and
replication of reported scientific results is a widely discussed topic
within the scientific
community~\cite{schwab-et-al:2000,barnes:2010,mesirov:2010,morin-et-al:2012,joppa-et-al:2013},
even spawning legislation in the USA~\cite{hr4012:2014}.  Whilst the
increasing number of high-profile retractions of scientific studies,
from climate science to bioscience, has drawn the focus of many
commentators, automated systems, which allow easy reproduction of
results, offer the potential to improve the efficiency of scientific
exploration and drive the adoption of new techniques. Nevertheless,
this is a wider socio-technical problem that pervades the scientific
community, with estimates that as much as 50\% of published studies,
even those in top-tier academic journals, cannot be repeated with the
same conclusions by an industrial
lab~\cite{osherovich:2011}. Furthermore, just publishing (linked and
open) scientific data is not enough to ensure the required
reusability~\cite{bechhofer-et-al:2013}. There are numerous
non-technical impediments to making software maintainable and
re-useable. The pressure to ``make the discovery'' and publish quickly
disincentivises careful software curation, with only recent
imperatives from funding bodies and governments trying to change this
position. Releasing code prematurely was often seen to give your
competitors an advantage, but we should be shining light into these
``black boxes''~\cite{morin-et-al:2012}. In essence: better software,
better research~\cite{goble:2014}.

We can thus exploit a fundamental advantage of the wider impact of
computer science and computational techniques to research: the unique
ability to share the raw outputs of their work as software and
datafiles. New experiments, simulations, models, benchmarks, even
proofs cannot be done without software. And this software does not
consist of simple hack-together, use-once, throw-away scripts;
scientific software repositories contain thousands, perhaps millions,
of lines of code and they increasingly need to be actively supported
and maintained. More importantly, with reproducibility being a
fundamental tenet of science, they should be re-useable. However, if
we closely analyse the scientific literature related to software tools
it often does not appear to be adhering to these
rules~\cite{nature:2011}. How many of them are reproducible? How many
explain their experimental methodologies, in particular the basis for
their benchmarking? In particular, can we (re)build the
code~\cite{collberg-et-al:2014}\footnote{In turn, stimulating further
discussions in this space:
\url{http://cs.brown.edu/~sk/Memos/Examining-Reproducibility/}}? We,
the authors, are perhaps as guilty as anyone in the past: we have
published papers~\cite{crick-et-al:2009a,berdine-et-al:2011} with
in-depth benchmarks and promises of code to be released online in the
near future.

However, we recognise that in many cases we are ``preaching to the
converted'' and do not need to sell the idea of reproducibility to the
computational science research community too much: we are used to
writing papers about our algorithms, models and tools; and reproducing
others' work (or having our own work reproduced) is encapsulated in
the ``Benchmark Tables'' and presentation of empirical results that
authors and referees in this community both think are essential to any
paper. The idea of reproducibility is gaining momentum across the
wider scientific research community, for example in computer
science~\cite{vitek+kalibera:2011,orchard+rice:2014,collberg-et-al:2015},
engineering~\cite{ledet-et-al:2014}, life
sciences~\cite{rollins-et-al:2014}, biomedical
sciences~\cite{huang+gottardo:2013}, climate
science~\cite{santer-et-al:2011}, ecology~\cite{thiele+grimm:2015},
epidemiology~\cite{peng-et-al:2006},
psychology~\cite{chambers-et-al:2014},
econometrics~\cite{koenker+zeileis:2009} and the social
sciences~\cite{conte-et-al:2012,hutton+henderson:2015}, as well as the
wide utility of creating and sharing reusable scientific workflows and
web
services~\cite{davidson+freire:2008,crick-et-al:2009b,oabarriaga-et-al:2014}. While
there has been a revolution in the sharing and dissemination of
published papers (\emph{open access}) and the subsequent discussions
relating to the sharing of protocols and materials (\emph{open
science})~\cite{pantonprinciples:2010,rssaaoe:2012}, the ability of a
researcher to {\emph{(a)}} keep track of the increasing number of
research outputs in their domain~\cite{parolo-et-al:2015} and
{\emph{(b)}} take these published results and data and reimplement the
described workflow remains
difficult~\cite{peng:2011,koop-et-al:2011,sandve-et-al:2013,wilson-et-al:2014}.

There has been promising work in this area, particularly around
benchmarking and
cyberinfrastructure~\cite{sim-et-al:2003,chirigati-et-al:2013,stodden+miguez:2014,stodden-et-al:2015},
along with a number of manifestos and community initiatives to
encourage and support reproducible research, such as the Recomputation
Manifesto~\cite{gent:2013}\footnote{\url{http://www.recomputation.org/}}
and cTuning~\cite{fursin-et-al:2014}, as well as curated
recommendations on where to publish research
software\footnote{\url{http://www.software.ac.uk/resources/guides/which-journals-should-i-publish-my-software}}. We
have previously encapsulated some of the technical barriers to
reproducing work across computing and the computational sciences,
particular in terms of the sharing of algorithms, models and benchmark
sets~\cite{crick-et-al_wssspe2,crick-et-al_recomp2014,crick-et-al:2015},
as well as drawing attention to some of the wider socio-cultural
issues~\cite{chuehong-et-al:2015}.

For example, although reproducibility is not a new concept to the
computer science research community~\cite{price:1986}, more recently a
number of high-profile computer science conferences, including PLDI,
POPL, SIGMOD, CGO, SPLASH and ECAI, have explicitly acknowledged the
importance of reproducibility\footnote{See
\url{http://www.artifact-eval.org/} and
\url{http://ctuning.org/reproducibility}} (and repeatability,
recomputability and the multitude of `Rs' that underpin
e-research~\cite{deroure:2010,feitelson:2015}), as well as promoting
community-driven reviewing and validation~\cite{fursin+dubach:2014}.
For many, this takes the form of the author providing \emph{artefacts}
--- an accessible tool for reproducing results --- for the reviewers
to evaluate. Journals such as Science, Nature, PLoS and from Royal
Society Publishing explicitly require that source code and data is
made available online under some form of open source license. While
these initiative are great, they are often optional, seem piecemeal,
and do little to easily enable verification or validation of
scientific results at a later stage. Even within the same field, there
are different ideas of what defines reproducibility.

In contrast to more traditional research outputs, the development of
algorithms has unique advantages. Algorithms are tested through their
implementation in code, and as such they can be accurately
communicated either by sharing the implementation and a pseudo-code
description of the algorithm behaviour. Furthermore, because of their
discrete nature it is expected that they aways return the same output
for the same input. It follows that these features are more easily
tested than the outputs of other scientific disciplines. To
demonstrate the reproducibility of a newly developed algorithm, one
must {\emph{(a)}} be able to create an operational artefact from the
code, and {\emph{(b)}} show that published models (i.e. benchmarks)
show the same behaviours as those reported. This specification for
reproducibility can in principle be tested for every publication in
the field.

Therefore, this paper invites computational researchers to embrace a
new methodology (and culture) for disseminating research. We propose an initial
specification for what a reproducibility service for key conferences and
journals in a field should look like. We present the requirements of the prototype,
and a suggested plan for introducing the service specifically to a high-profile
conference. In discussing the service, we highlight
key implementation issues relating to security and general
applicability which will need mitigating or resolving before
widespread acceptance by the research community.  The benefits of a
reproducibility testing service are clear. In a sense,
reproducibility here is an coming together of the standard practice of
\emph{testing} and ongoing work done in many continuous integration
systems (notably, automated build services). Three features are
prominent in our aims for reproducibility here: compilation and
testing in a new machine, a continuous integration strategy for code
commits and the ability to add and remove benchmarks to the test set.

We use a running example based upon
{\emph{BioModelAnalyzer}}\footnote{Available to use online:
\url{http://biomodelanalyzer.research.microsoft.com/}}, a tool for the
development and analysis (simulation, model checking) of a specific
class of formal models for biology, which has been described by the
authors over a series of VMCAI and CAV
papers~\cite{cook-et-al:2011,benque-et-al:2012,cook-et-al:2014}.  The
tool specifically allows users to test for model \emph{stability};
that is, a bespoke algorithm proves that for all initial states a
model always ends in a single, unique fixpoint. We have chosen this
example due to our familiarity with the tool, and to highlight
historical examples where a reproducibility service would have
supported both toolchain development and algorithm discovery.

\section{A Specification for Reproducible Computational Science}\label{spec}

A service for reproducibility is intended to play three important
roles. It should:

\begin{enumerate}[(i)]
\item Demonstrate that the source of an algorithm or tool can be
  compiled, run and behave as described, without manual intervention
  from the author; 
\item Allow new benchmarks to be added, by users other than
the developer, to widen the testing and identify potential bugs; 
\item Store and link specific artefacts with their linked
publications or other publicly-accessible datasets. 
\end{enumerate}

\subsection{{\emph{There Are Two Types of People}}}

\begin{figure*}[!htp]
	\centering
	\includegraphics[width=\textwidth]{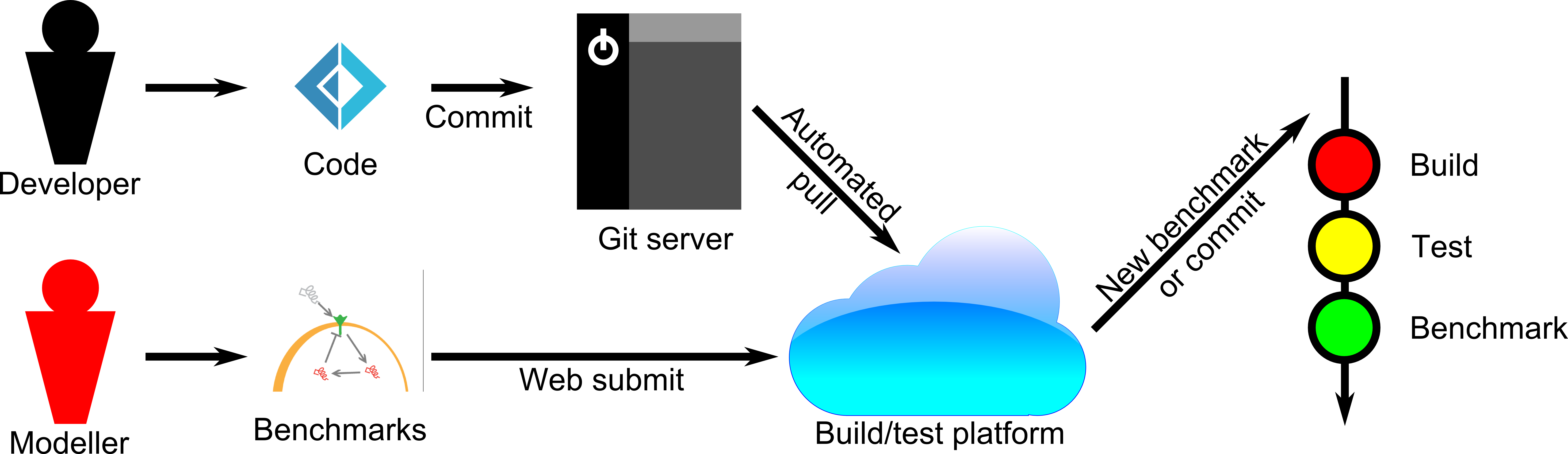}
	\caption{Proposed reproducibility service workflow}
	\label{schematic}
\end{figure*}

To address these needs, we propose that the service should follow the
workflow presented in Figure~\ref{schematic}. Two main classes of user are
defined; developers, who generate code; and modellers, who generate
new benchmarks. An individual might in practice play either or both of
these roles; here the roles serve to define ways in which people
interact with the system. A developer writes new code, which is
periodically pushed to a public repository such as GitHub. Through
integration with the repository, the server responds to new code by
undergoing a process of pulling the code from the repository,
downloading required dependencies, and compiling the code. If this
stage fails, the developer is informed and the workflow ends. If the
code is successfully compiled, two stages of testing are
performed.

The first stage (labelled {\emph{Test}} in Figure~\ref{schematic}),
involves running a series of basic tests defined by the
developer. This is intended as a sanity check to ensure that basic
features of the code have not been broken by the updated code, and
failure to pass these tests is reported and ends the workflow. If this
completes successfully, the second stage (labelled {\emph{Benchmark}}
in Figure~\ref{schematic}) runs and a series of models are tested for
a known property, and the results recorded. These results can then be
stored in a database, with a note of the commit ID, and available
through a web interface for future audit and analysis.

The service must fit easily into the developer workflow; as noted in
Section~\ref{rollout} we expect that there will be some costs to the
users in terms of the time required to ensure that the code compiles
and runs on the service. To minimise this, the service needs to
connect to standard code repositories, automatically detecting and
responding to new versions of the code and updates to dependencies,
running tests for every new code commit.

\subsection{{\emph{de Novo}} Build Environments}

To address the socio-cultural issues discussed earlier, a
service such as this must require minimal developer intervention.
This serves multiple purposes -- through automation for example, the
service can be enabled to compile new code and test new benchmarks
trivially. This also forces the developer to make publicly available
their local workarounds (i.e. hacks and workflow ``glue''). As such, this requires the
developer to make the project dependencies clearly available, and
enables future changes in the dependencies (such as a library update)
to be tested automatically too.

Throughout the lifetime of the {\emph{BioModelAnalyzer}} tool,
development has been shared between a number of developers, each
working on different aspects of the tool. Work in algorithm
development focuses on adding new features to a command line tool with
few dependencies, aiding rapid development. In contrast, the graphical
model construction and testing environment has typically been done by
a single or pair of individuals.  This necessarily required a number
of dependencies, reflecting the use of Azure (Microsoft's cloud
computing platform) and Silverlight (a framework for rich Internet
applications).

In an early stage of development it was found that only a single
machine was capable of deploying the web service. This arose as the
developer responsible for writing and deploying the user interface had
run a series of commands necessary to run the mixture of 32- and 64-
bit components on Azure. These commands needed only be run once, and
went undocumented, thus needing to be rediscovered later when other
team members attempted to deploy. These problems would be identified
trivially through the proposed service; such undocumented commands
would lead to all tests failing until explicitly added to the build
process.

\subsection{Tool refinement}

In contrast to the developer role, a modeller supplies benchmarks for
a piece of code to test against. These do not require that the latest
version of the code is recompiled, but on submission the models are
tested and added to the local repository of models for analysis.

In the case of {\emph{BioModelAnalyzer}}, throughout the development of the
tool, many refinements have been made to different
implementations. Some of these were subtle, and were identified by
unit tests; for example rounding mechanisms were switched between
floors and rounds following a scientific discussion. More complex
changes however broke behaviours which were not tested in our
available benchmark set. One example was in the treatments of nodes
without inputs; ``biological intuition'' suggested that such nodes
should have an alternative default function from other nodes. Here,
the ability of users to submit new benchmarks would aid identification
of these breaking changes, by extending the test sets and simplifying
the process of adding to the test sets, and forcing the question of
what changes are appropriate (and how to update old models to keep
correct behaviour).

\subsection{Identification of Algorithmic Weaknesses}

After a model is submitted, it is tested on every new piece of code
pushed to the server and the changes in the behaviour can be noted and
linked to specific code commits. Whilst the developer's role has a
transparent value (in providing an implementation of an algorithm),
the value of the modeller may be less immediately clear. The modeller
submits a broad range of tests which may highlight material flaws
(i.e. bugs) in the implementation, or the algorithm. More than this
however, the modeller may generate models which identify weaknesses of
either an algorithm or a specific implementation.

One example from the authors experience is the series of models with
``timed-switches'' described in~\cite{cook-et-al:2014}. There, we
presented a new algorithm for proving stability in a new class of
models. Whilst the paper focused on discussing the algorithm,
identifying the new class of models was complex. Models with long
cycles and the new class of models (``non-trivially stable models'')
both can take substantial time to search for cycles, and these models
could only be proved stable using a combination of simulation (to
identify the fixpoint) and LTL queries (to prove that there existed
no paths beyond a certain length which did not include the
fixpoint)~\cite{claessen-et-al:2013}.

\subsection{Algorithm-Model Axes}

The proposed service would allow both algorithm and model type of
tests to be included explicitly, and models to be routinely added to
each algorithm. Models which time-out with one but are successfully
proved can be logged and identified for future study. The features
which define them could then be more easily researched, and new algorithms
developed to address the specific features of the model. It could
further be used to demonstrate the speed improvement arising from new
algorithms.

\subsection{{\texttt{make depend}}}

Dependencies for a given implementation need explicit testing. Due to
the highly variable and sometimes complex nature of dependencies, we
see this as an optional part of the workflow, as developers may chose
to supply certain dependencies as binary files in the code compilation
process. For completeness however, we note that such a system could
also respond to updates in external dependencies by triggering
compilation and testing in the same manner as defined for a new code
commit. This would aid developers in identifying code breaking changes
introduced by third parties.

\subsection{``{\emph{I'm First!}}''}

Another issue is around performance comparisons of benchmarks: how can
we estimate, compare and evaluate raw performance in the cloud? Testing new
algorithms on benchmarks is in the first instance about pass/fail, but
very soon the focus is on raw performance. Benchmark tables are about out-performing other
algorithms, other tools. But, if the whole verification workflow is
running on the cloud, then acquiring and evaluating raw performance numbers is not
immediately feasible. There is no cloud equivalent of {\texttt{top}} or
{\texttt{time}} that gives user--resource statistics. There is too
much infrastructure interference/dependence --- with VMs spinning up, being torn
down, migrating, the bus being used by other VMs, etc --- to obtain
faithful numbers for the user/process/VM itself. Projects such
as Recomputation.org\footnote{\url{http://recomputation.org/}} have
been focusing on using virtual machines in the cloud to freeze, and
later unfreeze, computational experiments; while this approach is not
the complete solution, it is certainly a move in the right
direction~\cite{arabas-et-al:2014}. Nevertheless, the project's
primary aim is to validate recomputation~\cite{gent:2013}, with
performance a secondary consideration~\cite{gent+kotthoff:2014}, thus
making this a key avenue for further investigation.

\subsection{Running Arbitrary Code}

There are clearly significant potential security concerns around
providing open cyberinfrastructure that pulls, compiles and runs
arbitrary code as part of an autonomous continuous integration
framework; we need to consider precisely how this infrastructure would
interact with other open services, as well as privileges it would
require to run as an autonomous cloud service. Nevertheless, there are
existing models of sandboxing and privilege restriction from elastic
cloud computation providers that could be further developed and
applied.

\section{A Reproducibility Workflow for a Computational Research Community}
\label{rollout}

Following the proposal of such a system, the question becomes:
{\emph{how do we encourage widespread uptake, or even
    standardisation?}}  Such a service may appear non-trivial, given
the large numbers of tools and workflows that could potentially
require to be supported by the service. Furthermore, after such a
service has been implemented, how do we ensure it is \emph{useful} and
\emph{usable} for researchers. To address this, we propose the
following workflow that could be adapted and used for conferences in computational
science community:

\begin{enumerate}
\item {\textbf{Pre-conference:}} clear signposting for authors; it
should be advertised and promoted in the call for papers to
highlight this is a step-change in how we address
reproducibility. Call for artefact reviewers with a range of
specialisms, with a named chair of the review team.
\item {\textbf{Explicit criteria for authors:}} {\emph{make this as
easy as possible for us to evaluate/execute your artefact!}}. We would
aim to articulate the review criteria, but the primary aim is:
{\emph{can I evaluate/execute this artefact and get the same results
that are presented in the paper?}}
\item {\textbf{Submission:}} when papers are submitted, they have to
nominate whether they want their paper to go through artefact review
(at the start, this may not be compulsory, but this will change over a
period of time -- effecting cultural change and this would then become
a necessary condition and a formal stage in the reviewing workflow),
along with required tools, libraries and (ideally) computational
requirements.
\item {\textbf{Reviewing:}} in the first instance, it may be seen as
an extra (voluntary) step to the normal reviewing process:
e.g. {\emph{This submission is voluntary and will not influence the
final decision regarding the papers.}}. Independent of the scientific
merit of the paper, the results will be verified. To encourage this,
there may be a prize, as well as ranked ordering and profiled listed
in conference proceedings/final publication.
\item {\textbf{Artefact evaluation:}} artefact evaluation process runs
concurrent to the standard paper review process.
\item {\textbf{Reporting:}} traffic lights system (potentially with
ranked list) to indicate the level of reproducibility of the submitted
artefact.
\item {\textbf{Community curation:}} over a number of conference
  cycles, we would have a community curated repository/database of
  previous artefacts, which would provide exemplars, comparisons and
  emerging best practice.
\end{enumerate}

The key question for different research communities then becomes:
{\emph{how to initiate and initialise this change?}} Such a
requirement creates a set of new costs to researchers, both in terms
of time spent ensuring that their tools work on the centralised system
(in addition to their local implementation), but also potentially in
terms of equipment (in terms of running the system). Such costs may be
easier to bear for some researchers compared to others, especially those
with large research groups who can more easily distribute the tasks,
and it is important that the service does not present a barrier to
early career researchers and those with efficient budgets (this type
of cost analysis is not unique to reproducibility efforts -- it has
been estimated that a shift to becoming exclusively open access for a
journal may lead to a ten-fold increase in computer science
publication costs~\cite{vardi-cacm-2014}). Another advantage of such a
service however is that it sets a clear minimum standard for
reproducibility in computational research. It does not impose a specific set
of licences (a limited licence for testing would be all that is
necessary). By its nature, it can be uniformly applied to all users in
a conference, preventing the ``weaponisation of reproducibility'' that
has been highlighted as a potential
weakness\footnote{\url{http://simplystatistics.org/2015/03/13/de-weaponizing-reproducibility/}}.

\section{Conclusions}\label{concl}

The benefits to the community from a cultural change to foster and
favour reproducibility are clear and as such we should aim through the
development and adoption of open cyberinfrastructure, software
toolchains and workflow to mitigate these costs. Furthermore, we can
reasonably expect the needs of the community to evolve over time, and
initial implementations of the platform may require refinement in
response to user feedback. As such, if the community is to move to
requiring reproducibility, it seems most reasonable that this is
staggered over a number of years to allow for both of these elements
to develop, until eventually all researchers are required to use the
service. This plan balances competing needs within the community, and
would reduce the disruption for uptake by gradually introducing it to
researchers.

\begin{itemize}
\item {\textbf{Year {\emph{t}}:}} Offer the service as an optional extra in the
  testing phase, allowing users to demonstrate the reliability of
  their code which could be taken into account in the review process.
\item {\textbf{Year {\emph{t+1}}:}} All authors must use the reproducibility
  service, but results are not used in the review process. The results
  of the test are used to refine the service and pick out any
  unaddressed issues
\item {\textbf{Year {\emph{t+2}}:}} All authors are required to use the service, and
  the results are explicitly used to assess reproducibility in the
  review process.
\end{itemize}

This open discussion, understanding and acceptance of what
reproducibility means for the wider computational research community
is clearly important. It is imperative that we see it as worthwhile
and address it, but we now have to move beyond manifestos, pledges and
top tips: as researchers, we all need to publicly acknowledge that
this is worthwhile and put concrete measures in place to address it,
or stop going on about it.



\bibliographystyle{IEEEtran}
\bibliography{cse2015}

\end{document}